%% file: nanostreams_submission.tex
\pdfoutput=1
\documentclass[11pt]{article}

\usepackage{color}
\usepackage{todonotes}
\usepackage{verbatim}
\usepackage{booktabs}
\usepackage{graphicx}
\usepackage[colorlinks=true,linkcolor=black,citecolor=black]{hyperref}
\usepackage{epstopdf}
\usepackage[toc, page]{appendix}
\usepackage{subfigure}
\usepackage{chngcntr}

\usepackage[margin=2cm]{geometry}
\usepackage{palatino}
\usepackage{enumitem}

\def\affilnum#1{${}^{#1}$}
\def\affil#1{${}^{#1}$}
\def\address#1{{\footnotesize\itshape #1}}
\def\corraddr#1{\footnotetext[1]{Correspondence to: #1\stepcounter{footnote}}}
\def\keywords#1{\small{KEY WORDS:}{#1}}
\def\corrauth{\footnotemark[1]}
\RequirePackage{graphicx}
\RequirePackage{pifont,latexsym,ifthen,rotating,calc,textcase,booktabs,color}
\RequirePackage{amsfonts,amssymb,amsbsy,amsmath,amsthm}
\RequirePackage[errorshow]{tracefnt}

\title{Iso-Quality of Service: Fairly Ranking \\
Servers for Real-Time Data Analytics}
\begin{document}
\typeout{  >>>> Start processing document }
\date{}
\author{Giorgis Georgakoudis\affil{1}\corrauth,\ 
        Charles J. Gillan\affil{2},
        Ahmed Sayed\affil{1},\\
        Ivor Spence\affil{1},
        Richard Faloon\affil{3},
        and Dimitrios S. Nikolopoulos\affil{1}}
\maketitle
   %

\begin{center}
\affilnum{1}The School of Electronics, Electrical Engineering and Computer Science,
                  Queen's University Belfast, Northern Ireland BT7 1NN, United Kingdom \break
\affilnum{2}The Institute for Electronics Communications and Information Technology,
                 Queen's University Belfast, The Northern Ireland Science Park
                  Queen's Road, Belfast, Northern Ireland BT3 9DT, United Kingdom  \break
\affilnum{3}Neueda Consulting Limited, Glenwood Business Centre, Springbank Industrial Estate
               Belfast, Northern Ireland BT17 0QL, United Kingdom
\end{center}

\corraddr{The School of Electronics, Electrical Engineering and Computer Science,
          Queen's University Belfast, Northern Ireland BT7 1NN, United Kingdom}
   %
\input{abstract}
   %
\input{introduction}
\input{options}
\input{setup}
\input{qos_analysis}

\input{discussion}
\input{related_work}
  %
\input{nanostreams_project}
  %
\typeout{>>>> Processing the Conclusions section}
\input{conclusions}
\begin{center}
\textbf{Acknowledgments}
\end{center}
The work was supported by the European Commission under its Seventh Framework Programme, grant
number 610509 (NanoStreams). This work was also supported by the UK Engineering and Physical Sciences 
Research Council, under grant agreements EP/K017594/1, EP/L000055/1 and EP/L004232/1.
\typeout{>>>> Creating the references now}
\bibliographystyle{wileyj}
\bibliography{nanostreams_submission}
\typeout{****  Finished creating the references now - End of document}
\end{document}

%% file: abstract.tex
%
%
   %
\begin{abstract}
We present a mathematically rigorous Quality-of-Service (QoS) metric which
relates the achievable quality of service metric (QoS) for a real-time analytics
service to the server energy cost of offering the service.
Using a new iso-QoS evaluation methodology, we scale server resources 
to meet QoS targets and directly rank the servers in terms of their energy-efficiency and by
extension cost of ownership.  Our metric and method are platform-independent and enable
fair comparison of datacenter compute servers with significant architectural 
diversity, including micro-servers.
   %
   %
We deploy our metric and methodology to compare three servers running financial option pricing workloads
on real-life market data. We find that server ranking is sensitive to data inputs and desired QoS level and
that although scale-out micro-servers can be up to two times more energy-efficient than conventional 
heavyweight servers for the same target QoS, they are still six times less energy efficient than high-performance computational accelerators.  
\par   
   %
   %
\end{abstract}
   %
%
%

%% file: introduction.tex
\section{Introduction}
Sustaining a defined Quality of Service (QoS) is an integral part of any 
Service Level Agreement (SLA)  pertaining to the provision of enterprise level
compute services. 
These compute services
run on large data centers. The key business driver for the owners of these centers is the profit to be made by charging end users for the  services provided. QoS provision is an integral part of the owner profit
and user cost model of datacenter and datacenter services. 
\par
Emerging services providing real-time data analytics, such as trade and credit risk analytics in the
capital markets, incur a high usage and hosting premium. The reason is that the computational
workloads of these services are highly dynamic, event-driven, and demanding in terms
of target real-time response latency, which is often measured in microseconds. QoS provisioning for 
such services requires significant investments in server and networking infrastructure, in addition
to painstaking optimization of the service software.
\par
A central question in provisioning hardware for real-time data analytics is the choice of compute server
architecture that will meet the latency targets of the service, while reducing the operational cost of the datacenter and energy consumption in particular. The choice is challenging because of vast differences between servers in architecture, price points, operational points, and target markets. As an example, the experimental campaign that we conducted for this paper suggests that a given QoS target for real-time option pricing workloads on actual market data feeds may be met by server hardware with power budgets ranging from 25W to over 200W and latencies ranging by a factor of five. How does the datacenter owner choose the best server for low-latency,
real-time analytics workloads? Conversely, how does a user select the best equipped datacenter to run the same class of workloads? This paper sets to address these questions. 
\par
In this paper we present a new QoS metric for the fair ranking of servers that support real-time analytics
workloads with low latency requirements. The metric allows direct comparison between servers in terms
of raw performance  and energy-efficiency, while equating the QoS that they provide to users. This
leads to an iso-QoS approach for ranking servers. We present a mathematically rigorous metric that
accurately models dynamic workloads with real-time event response deadlines and demonstrate that our 
metric fits well real-life financial option pricing workloads on actual market data. 
The metric and its derivation are platform-agnostic and can be used directly to optimize server
provisioning for energy cost minimization under SLAs.
\par
We mine data presented in previous papers~\cite{Georgakoudis2014,Gillan:2014:VMF:2688424.2688429}
to rank three servers in terms of iso-QoS under option pricing workloads: a scale-out microserver
based on Calxeda SoCs; a dual-socket Intel Sandy Bridge server; and an Intel Xeon Phi server. 
Our experimental campaign uses option pricing workloads for which we invested identical
effort to optimize on each server. The campaign reveals new findings: The scale-out microserver
can be up to two times more energy-efficient than heavyweight servers under iso-QoS, but six times
less energy-efficient than a high-performance co-processor. Importantly, the relative ranking of servers
varies with the option pricing algorithm and input to the algorithm, while changing server provisioning
produces also counter-intuitive rankings. 
\par
The paper begins by briefly defining financial option contracts and 
their use in our real-time workloads in Section~\ref{sec:option_pricing}.
We move on to details of the platforms used 
and a summary of our experimental methodology  
in Section~\ref{sec:Experimental Setup and Measurement Methodology}.
We present our mathematical model for QoS next, in Section ~\ref{sec:basis_of_the_qos_metric} and 
apply an iso-QoS for two option pricing kernels to rank platforms in terms of energy efficiency. 
In Section~\ref{sec:discussion} we discuss the results of our experimental campaign, while in Section 
~\ref{sec:related} we present related work in the field.
Section~\ref{sec:nanostreams}
describes the Nanostreams project within the context of which this work took place.
The paper is concluded in Section~\ref{sec:conclusions}.

%% file: options.tex
\section{Computing Option Prices}
\label{sec:option_pricing}
A financial Option is a contract giving the owner the right 
to either sell (Put) or to buy (Call) a fixed number of assets, frequently company stock,  
for a defined price on, (European option) or before (American option) an end date. 
Methods from stochastic calculus produce equations to model 
option prices by simulating multiple paths of the underlying variables 
over a time window. Analytical solutions for these equations 
are not generally possible so a variety of  computational numerical 
solution methods have been developed. We construct real-time analytics 
workloads that continuously execute Monte Carlo (MC) or Binomial 
Tree (BT) option pricing models. 
\par
European vanilla options are a particular subset of option types.
Black and Scholes~\cite{blackscholes:1972,blackscholes:1973} proposed a 
second-order partial differential equation which models the variation of 
an option price with contractual strike price $P$, 
over time $T$ years to contract expiry, assuming that the underlying asset 
spot price, 
$S$ follows a log normal distribution and that the volatility $\sigma$ of $S$ 
the risk free rate of return, $r$, are constant. An analytic solution to 
this equation exists 
for European vanilla options but not generally for other types of options. 
Our work focuses on European vanilla options because we can then use the 
Black-Scholes solution to provide a reference against which to compare 
our code base and its generated numerical results for accuracy.
\par
A rich literature already exists for both the MC and BT methods\cite{boyle:montecarlo,coxrossrubenstein:1979}.
therefore we present them only briefly here. 
An MC simulation computes the current price of a Put contract by
\begin{equation}
   {\rm Price} = 
   \frac{ {\rm e}^{-rT} }{N} \sum_{i=1}^{N} 
       \max \left  (
                    0, S - P {\rm e}^{ 
                                      ( r - \frac{\sigma^2}{2} ) T + \sigma \sqrt{T} x_i 
                                     }
            \right ) 
   \label{eqn:montecarlo}                                            
\end{equation}
where ${x_i\ (i=1\ldots N)}$ is a set of random numbers drawn from the 
standard normal distribution. We generate these using the 32-bit version 
of the Mersenne Twister algorithm~\cite{mersenne:32} and the Box-Muller 
transformation.
The BT pricing model discretises the time to expiry, $T$ in years, into a lattice of $N+1$
levels with the root node as the current underlying asset price $S$. Starting at the 
root, an up and a down factor are applied to generate two prices at the 
next level. This continues, using the same constant factors, for all prices 
at all levels until the end level is reached. The final stage of the 
algorithm works backwards over the lattice computing an expectation value for
each price at each level, finishing at the root node, which then contains the current 
option price.
\par
Both algorithms depend on a parameter $N$ and both converge non-monotonically 
to an exact answer in the limit $N\rightarrow\infty$. However they have different 
computational characteristics. Generic MC is a classic ``for" loop summation, requiring evaluation
of transcendental functions, and its operation count scales as $O(N)$ while the 
BT is dominated by a nested for-loop of add-multiply operations implementing 
the backward propagation step and scaling as $O(N^2)$.

%% file: setup.tex
\section{Experimental Setup and Measurement Methodology}   
\label{sec:Experimental Setup and Measurement Methodology}
Our experimental setup includes three platforms on which we execute our
OptionPricer program and collect workload-specific performance and 
energy metrics. 
This Section defines our metrics, 
describes the platforms used and presents salient details of our methodology
used to obtain the power readings and calculate the energy consumption.
A complete description of our methodology is available  
in \cite{Georgakoudis2014}.

\subsection{Definition of Metrics} \label{sec:platforms:metrics}
Option pricing in finance takes place by consuming 
a live streaming data feed of stock market prices, often within the 
context of high frequency trading (HFT), and for pre-trade risk analytics.
The execution time characteristics of option pricing are different 
from those of numerical simulation in computational science using HPC.
By contrast to scientific codes which have measurable setup and 
post-processing phases, financial option pricing runs relatively small 
standalone kernels, such as MC and BT, at very high frequency with 
little set up and post processing work.
Option pricing on live market data feeds is actually a form of event 
processing, where the event is the arrival of a price update on the 
underlying stock. Based on these distinctions we present and use three 
workload-specific metrics to compare servers under financial analytics 
workloads:
\begin{description}[leftmargin=0cm]
  \item[QoS] New prices may arrive at any time in a trading 
             session. This means that any contracts not yet 
             priced using the previous price update are 
             abandoned and deemed unusable. Related to the 
             Time/option metric below, 
             but also dependent on market activity, we define 
             the Quality of Service metric (QoS) as the ratio 
             of successful to the total requested option 
             price evaluations. The QoS metric is an 
             application-specific measure on meeting 
             option pricing performance requirements. It is 
             useful for characterizing application-related 
             performance and scalability offered by deploying 
             multiple nodes. It is worth noting that QoS 
             depends on the rate of stock price changes and other 
             market activities at the time of its calculation,
             so it will be different each time it is calculated 
             in a live market scenario.
  \item[Joules/option] (J/Opt or J$_{\rm opt}$) The energy consumed per execution of a pricing kernel is
         a fundamental metric.
In the case of 
         an actively traded stock, with a high 
         number of defined option contracts, this building
         block is executed repeatedly throughout the trading day.
         Correspondingly, a reduction in this value can result in significant
         energy savings for providers offering option pricing services.
 \item[Time/option] (S/Opt or S$_{\rm opt}$)
         In contrast to providers, end users, particularly those engaged in HFT, 
         are sensitive to end-to-end latency, thereby constraining the elapsed 
         time per option metric. This metric in turn can be used to evaluate the 
         total time to price all contracts for a given stock. Option pricing 
         shares this time-to-solution performance metric in common with HPC applications. 
\end{description}
\subsection{Hardware Platforms}  \label{sec:platforms:platforms}
We used three platforms, one state-of-the-art server architecture with Intel Sandy Bridge 
processors (briefly referred to as ``Intel'' in the rest of this paper),
one state-of-the-art 
HPC architecture with Intel Xeon Phi Knights Corner coprocessor 
(referred to as ``Xeon Phi'') and a Calxeda ECX-1000 microserver with 
ARM Cortex A9 processors, packaged in a Boston Viridis
rack-mounted unit (referred to as ``Viridis''). 
We used the $4.7.3$ version of the GCC compiler and the Intel Compiler ICC 
version $14.0.0 20130728$ 
for code generation, the latter only on Intel platforms.  
The three platforms offer the possibility of 
scaling their frequency and voltage through a DVFS interface.  
We conducted experiments only with the highest voltage-frequency 
settings on each 
platform, to which we refer as performance mode. Previous work shows 
that performance mode is the most energy efficient too~\cite{Georgakoudis2014}.
The details of the platforms are as follows:
\begin{description}[leftmargin=0cm]
\item[Intel] is an x86-64 server with Sandy Bridge architecture, 
             with 2 Intel Xeon CPU E5-2650 processors 
             operating at a frequency of 2.00GHz and equipped 
             with 8 cores each. The machine has 32GB of DRAM (4 $\times$ 8GB DDR3 @ 1600Mhz). 
             The server runs on Linux CentOS 6.5 with kernel 
             version $2.6.32$ ($2.6.32-431.17.1.el6.x86\_64$).
\item[Xeon Phi] (Knights Corner) is a many core, 
                x86-64 co-processor board (5110P model) over PCIe. 
                It features the many integrated cores (MIC) 
                architecture which offers sixty, 4-way hyperthreaded 
                cores, each equipped with a very wide (512-bit) vector 
                unit. The board has more than $6$~GB of GDDR5 DRAM. 
                and the clock frequency is $1.053$ GHz. 
                High performance and high energy efficiency are 
                the result of featuring a highly parallel many core 
                design while running in low clock speeds.
                The system runs on Linux kernel 2.6.38.8+mpss3.2.1.
\item[Viridis]   is a 2U rack mounted server containing sixteen  
                 microserver nodes connected internally by a high-speed 10 Gb Ethernet network. 
                 The platform appears logically as sixteen servers within one box. 
                 Each node is a Calxeda EnergyCore ECX-1000 comprising 4 ARM Cortex A9 cores and 4~GB of DRAM
                 running Ubuntu 12.04 LTS. 
                 Viridis has a frequency of $1.4$GHz.
\end{description}
Note, when referring to the different platform settings later we will use the 
following notation to represent the platform configuration 
[Nodes used $\times$ Cores Used $\times$ Threads per Core].
\subsection{Software}
Starting from a common C code base, we created versions which use the 
vector units on each platform. We achieved this in three different 
ways
\begin{itemize} 
   \item creating assembler code implementations of hotspot loops
   \item using compile intrinsic C functions which map to assembler
         instructions
   \item using the auto vectorization functionality of the kernel.  
\end{itemize}   
\begin{table}[tbhp]   
  \footnotesize
  \centering
  \caption{List of labels, VEC TYPE, defining the preparation of the executable binary}
  \begin{tabular}{ll} 
     \toprule
       \multicolumn{1}{c}{\bf VEC TYPE} & \multicolumn{1}{c}{\bf Description}        \\
     \midrule 
       \textbf{AVX256}     & Assembler code using AVX 256-bit instructions on the Intel Sandyridge. \\       
      \midrule       
       \textbf{INTRINSICS} & Compiler supplied C functions on any platform (ARM 128-bit, Intel 256-bit, Xeon Phi 512-bit) \\    
      \midrule       
       \textbf{KNC512}     & Assembler code for 512-bit vector instruction set on the Xeon Phi (Knights Corner). \\ 
      \midrule       
       \textbf{NEON128}    & Assembler code for the ARM NEON 128-bit unit. \\ 
      \midrule
       \textbf{AUTOVECT}   & Compiler auto-vectorization on all platforms \\
     \bottomrule
  \end{tabular}
   \label{tab:binary_labels_table} 
\end{table}  
Table \ref{tab:binary_labels_table} defines the labels corresponding to the 
type of binary. Each experiment, reported later in this paper, is conducted
by executing one type of binary on one platform and is labeled accordingly.
\subsection{Summary of Methodology}  \label{sec:platforms:methodology}
\begin{figure}[htbp]
  \centering
  \includegraphics[width=0.4\textwidth]{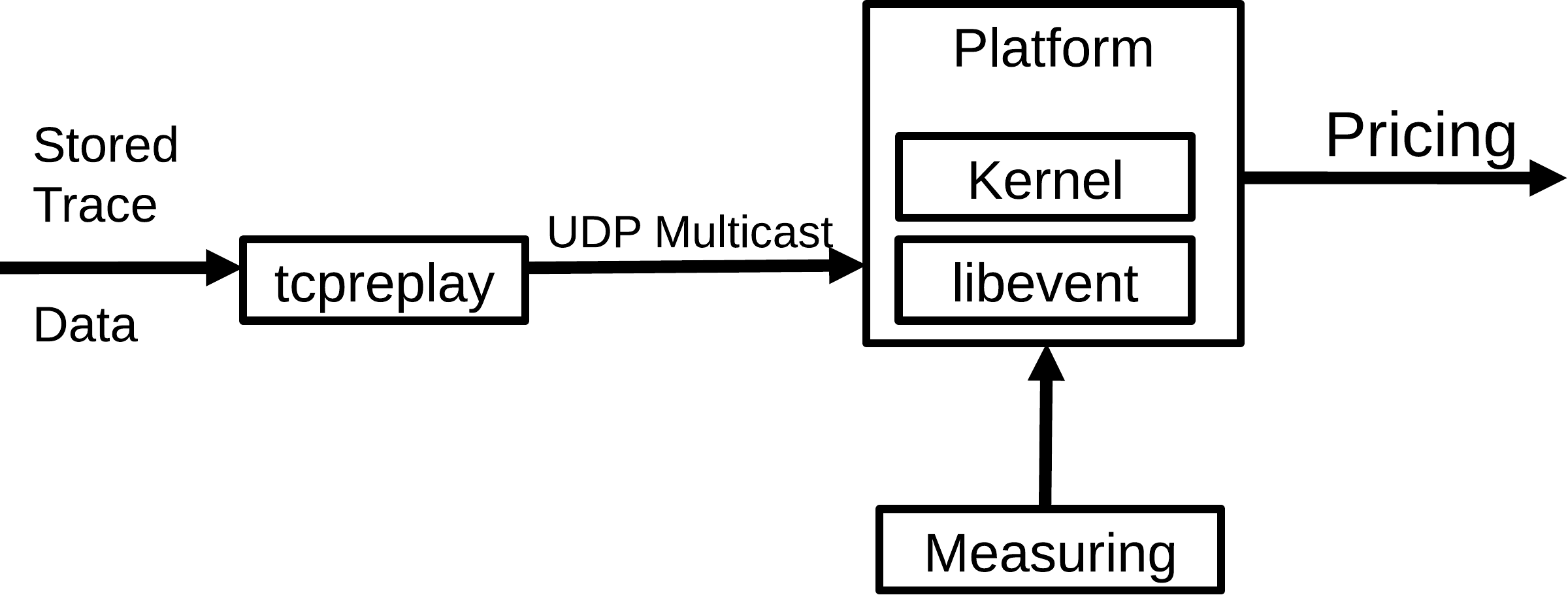}
  \caption{Financial trace data measurement setup}
 \label{fig:fig1}
\end{figure}

For our experiments, we collected Facebook stock price ticks during a 
full New York Stock Exchange session and replayed them using UDP 
multicast to all nodes in each of our platforms, as shown in 
Figure~\ref{fig:fig1}. This is as close as an experiment 
needs to be to reality without any external glitches or factors 
affecting the setup or measurements. 
Detection of a change in the Facebook stock price triggers computation 
of new prices for 617 Facebook European options at the maximum speed 
feasible.
\par
\begin{figure}[htbp]
  \centering
  \includegraphics[width=0.45\textwidth]{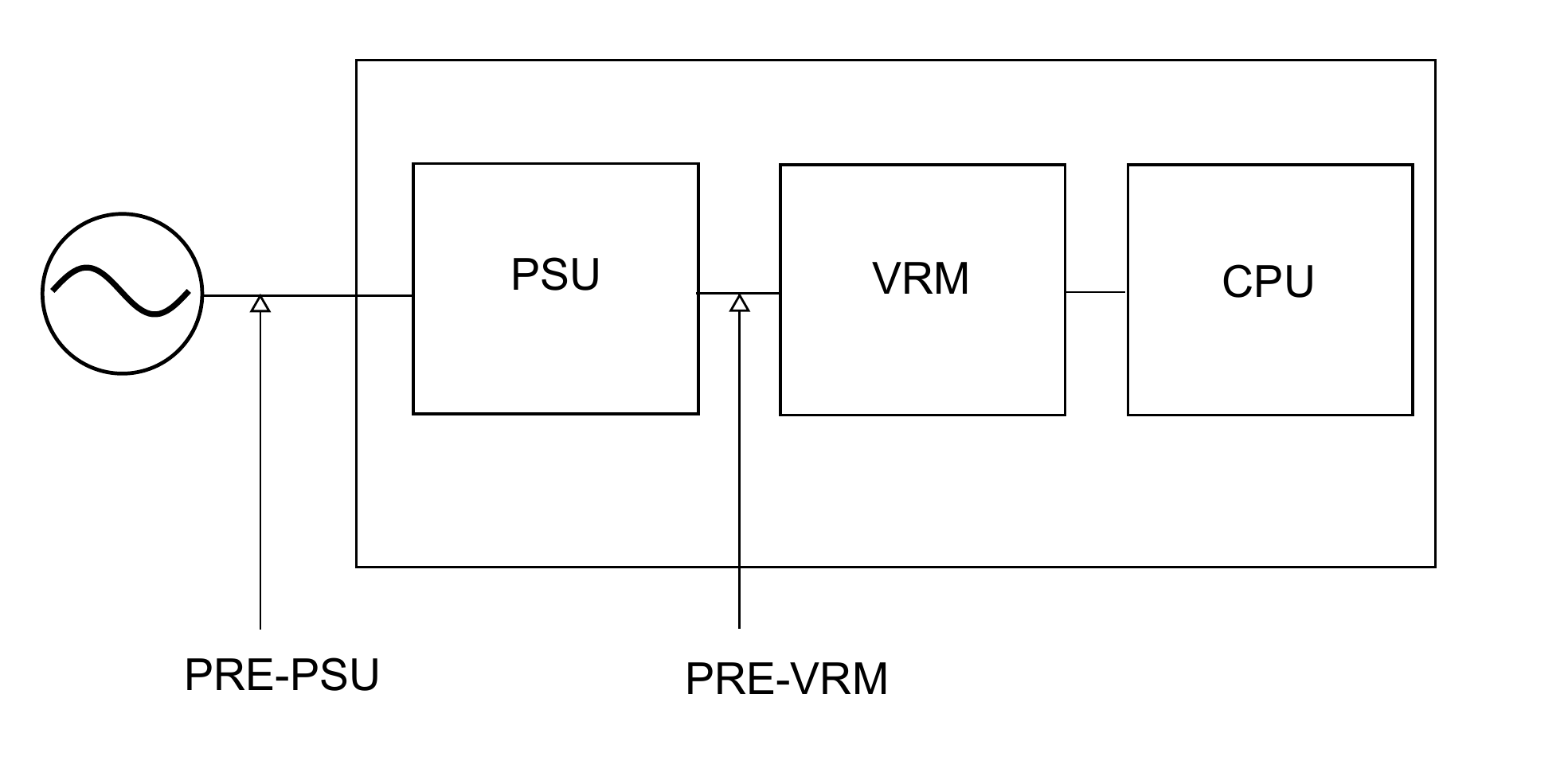}
  \caption{The path of the current supply to the CPU showing points at which
           we measured power. PSU is the power supply unit and VRM the 
           voltage regulator module.}
  \label{fig:currentsupplypath}
\end{figure}

Next we discuss on the power measurement methodology.
The exact form of the current supply path to the CPU differs from one 
platform to the next but to provide a fair basis for comparison 
we identified two distinct points on the path,
shown in Figure~\ref{fig:currentsupplypath}, which are measurable 
on all platforms. We continuously monitored power on each platform
at these points during our experiments. 
To isolate the energy consumption of processor packages, we capture 
power consumption at the point before the VRM, which we label PRE-VRM. 
For the Intel server, PRE-VRM measurement
is facilitated by reading the Running Average Power Limit (RAPL) 
counters while the same
functionality on Viridis is available through the Intelligent 
Platform Management Interface (IPMI)
counters, which is also available on the Xeon Phi platform
\par
\begin{figure}[htbp]
   \centering
   \includegraphics[width=0.48\textwidth]{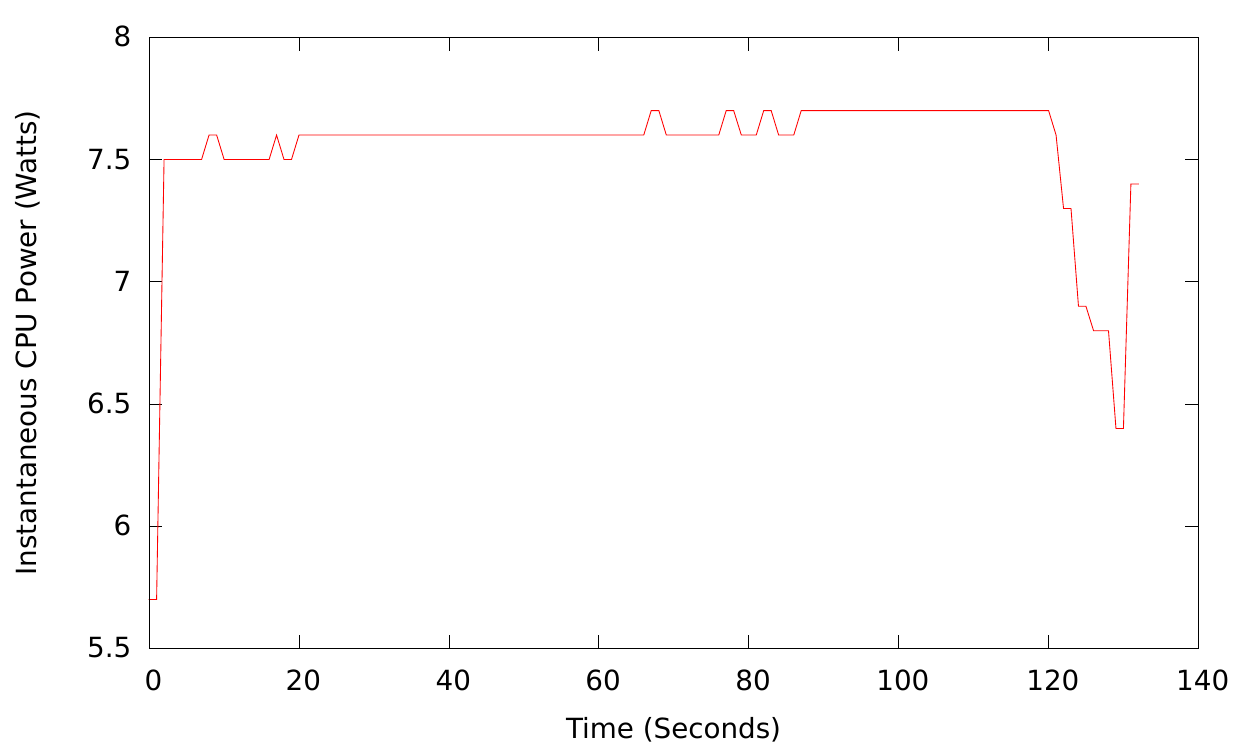}
   \caption{CPU power vs. time for the MC kernel}
   \label{fig:powervstime}
\end{figure}

Figure~\ref{fig:powervstime} shows the power versus time plot for a
standalone execution of the MC kernel. The BT execution plot is similar. 
The profile of instantaneous power versus time follows 
a very sharp trapezoidal shape: the CPU is fully utilized during execution and 
there are no periods of inactivity. This is a common feature with other 
numerically intensive HPC applications. It means that the measured 
average power is a representative measure of energy consumption 
throughout kernel execution. 

%% file: qos_analysis.tex
\section{The Mathematical basis of the QoS Metric}  \label{sec:basis_of_the_qos_metric}
Many of the worlds leading financial trading venues are order driven markets, 
meaning that investors, especially high frequency traders, submit buy and sell 
orders independently to matching engine software operating at high speed at the 
venue. These engines cross buy and sell orders to create trades and are a
key part of the electronic trading platforms which underpin high frequency 
trading. Sequential models, which are the basis to analyze trading 
patterns in high frequency trading, assume a Poisson distribution to 
model the arrival of orders affecting stock price into the system.
\subsection{The QoS as a cumulative frequency distribution}
In this section we explain how we create a QoS curve as a function of
price gap frequencies. It is important to note that this curve is dictated
solely by the market activity. In the next section we explain how we 
can determine using the S$_{\rm opt}$ ad J$_{\rm opt}$ metrics for a given
platform whether we can meet a required QoS value or not.
\par
From our data, we created a histogram of the distribution of time gaps 
between price 
updates for the Facebook stock and from this computed a cumulative frequency 
distribution (CFD) which we noted exhibits the characteristics of a Poisson 
CFD. This reflects the assumptions of the sequential model of financial 
trading. 
\par
Normally in a CFD the value assigned to bin $i$ is the sum of all values in 
bins $1,\ldots, i$. In our case these are time bins so that the frequency is 
the number of price updates arriving at time intervals up to and including 
that represented by bin $i$. There is a value of the time gap, depending
on the performance of the platform, the number of options to be priced and the 
kernel used, below which it is not possible to satisfy the hard constraint of 
computing prices for all defined options. We denoted this by $G$. Our QoS 
metric actually corresponds to the sum over all time bins greater than this 
threshold. It follows that our QoS function is obtained by reflecting the 
initial CFD around its mid-point on the time axis. This means that we can fit our 
observed time gap distribution to the form
\begin{equation}
  {\rm QoS(t)} = 1 - e^{-\lambda} \sum_{i=0}^t \frac{\lambda^t}{\lfloor t!\rfloor}
\end{equation}
Furthermore, we define 
the QoS, the y-axis, as a percentage rather an absolute value.
\par
\begin{figure}[htbp]
  \centering
  \includegraphics[width=0.45\textwidth]{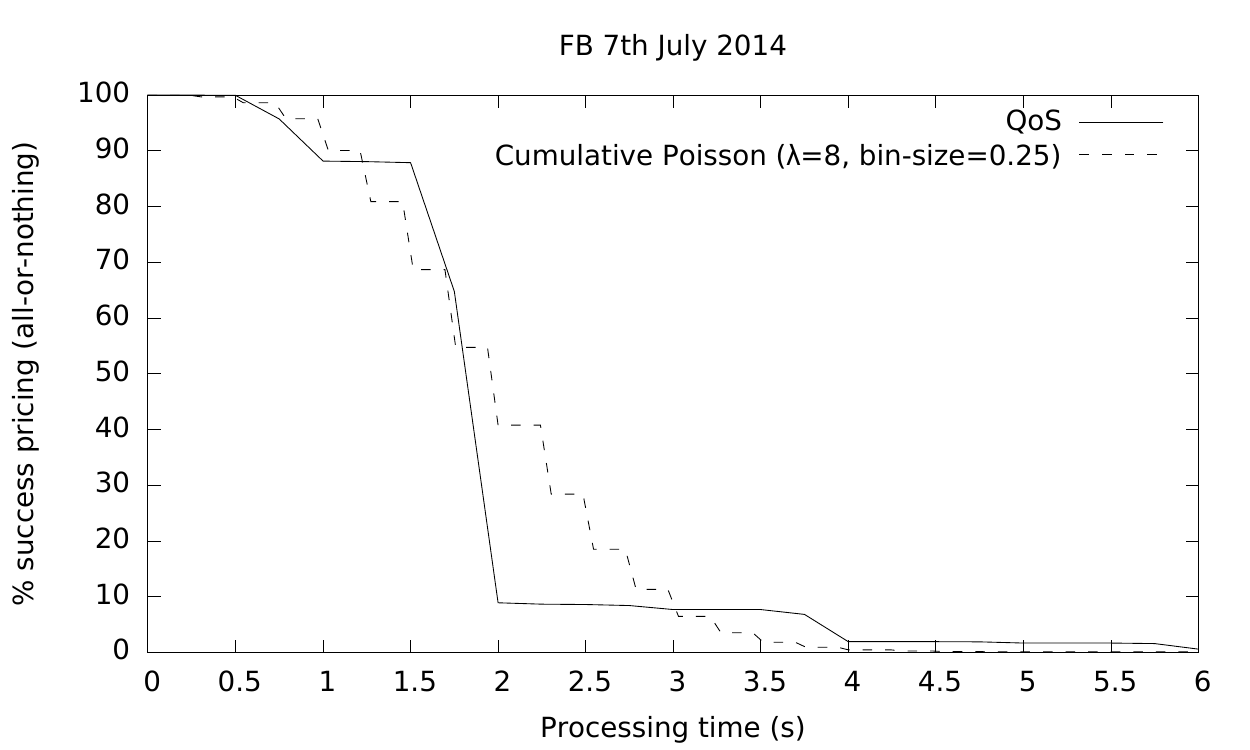}
  \includegraphics[width=0.45\textwidth]{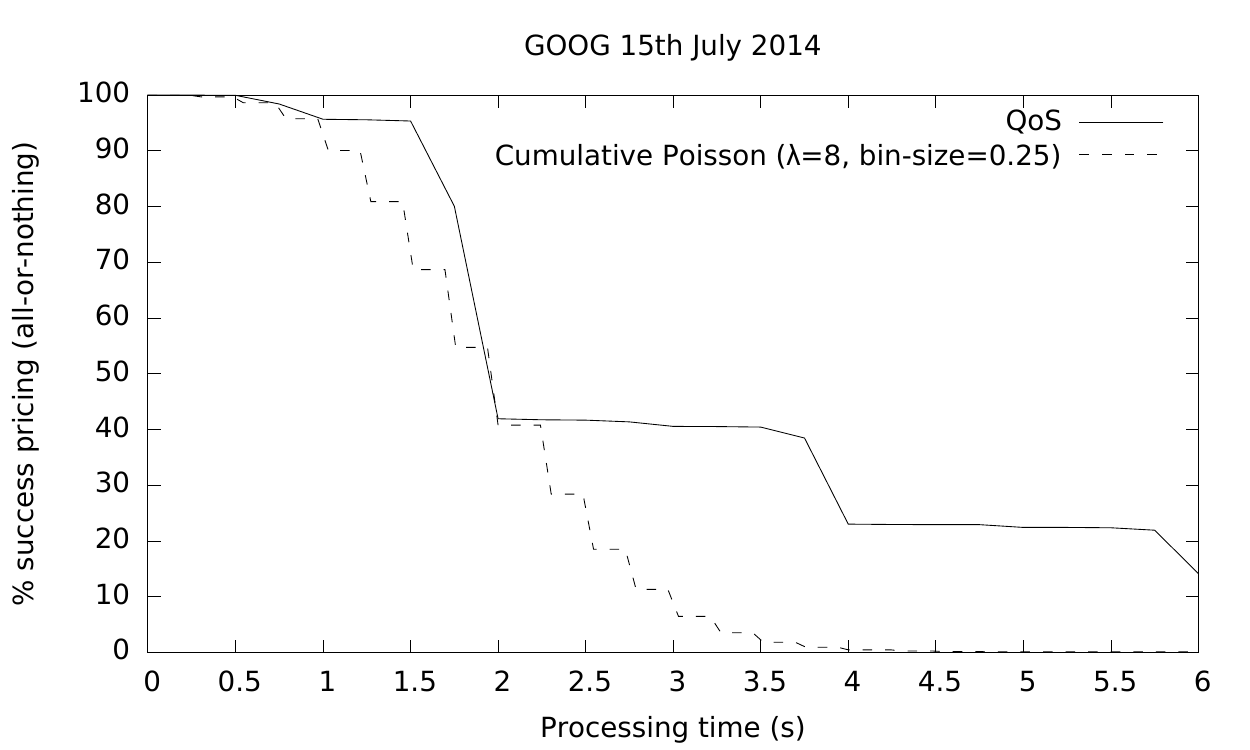}
  \caption{Cumulative frequency distribution of Facebook and Google stock price updates 
           for full trading sessions on July 7th and 15th 2014}
  \label{fig:fittedcumulativepricefreq}
\end{figure}

The data for our experiments are taken from a trading
session of 6.5 hours where $10,156$ price updates occurred 
for the Facebook (FB) stock, resulting in the cumulative 
distribution function representing the QoS shown in 
figure \ref{fig:fittedcumulativepricefreq}.
The solid line shows the measured values joined directly by straight lines
while the dashed curve shows the result of fitting the measured data
to the analytic expression for the cumulative Poisson distribution.
Further confirmation of the Poisson-like behavior of the arrival of price updates is seen in the 
profile for the Google stock which is also presented in figure \ref{fig:fittedcumulativepricefreq}.
Similar price update profiles occur in work \cite{DBLP:conf/icde/LiuWWZBR10} studying prices on the German DAX exchange. 
\subsection{iso-QoS and total energy consumed}   
Let us set a required QoS $Y\%$ for all our platforms. From the QoS curve
we can determine a minimum time constraint, $G$, that we must satisfy. 
Within $G$ seconds we need to compute all $N_{\rm opt}$ options defined 
on the stock. First of all a platform can only satisfy this constraint if 
\begin{equation}
     G \geq N_{\rm opt} \times S_{\rm opt} \label{eqn:constraint}
\end{equation}
Assuming this is met, we know that the energy consumed in each time gap
is then 
\begin{equation}
     E_{\rm gap} = N_{\rm opt} \times J_{\rm opt}
\end{equation}
where we ignore idle power. 
Next, we know from the definition of QoS 
that the total number of time gaps in which we will perform the 
computation is 
\begin{equation}
    N_{\rm gaps} = {\rm floor}(Y \times {\rm Total   \;\; number \;\; of  \;\;
                                             updates \;\; for    \;\; the \;\; 
                                             session)}
\end{equation}
so that the energy consumed doing option pricing while meeting QoS $Y\%$ is
\begin{equation}
   E_{\rm {QoS=Y}} = N_{\rm gaps} \; \times \; E_{\rm gap} 
\end{equation}
Platforms may then be ranked, for this QoS, in order of energy 
consumption.   
\subsection{Application to platforms}
We have applied the equations defined above using the QoS curve 
in figure \ref{fig:fittedcumulativepricefreq}. 
Table \ref{tab:mc_0.5M_qos_10_table} is the result of the analysis
of delivering option pricing with a $10\%$ QoS using the MC kernel 
operated with 0.5M iterations. Only the five cases (platform plus software) 
which can satisfy the constraint in equation (\ref{eqn:constraint}) 
are reported.
\begin{table}[htbp]
   \footnotesize
   \centering
   \caption{MC kernel (N=0.5M and QoS=10\%)}
   \begin{tabular}{lrrrrrrr} 
     \toprule
     \multicolumn{1}{l}{\bf Platform}   & 
     \multicolumn{1}{c}{\bf VEC TYPE}   & 
     \multicolumn{1}{c}{\bf S/Opt}      & 
     \multicolumn{1}{c}{\bf J/Opt}      &
     \multicolumn{1}{c}{\bf Energy(KJ)}    \\
     \midrule 
       \textbf{Viridis(16$\times$4$\times$1)}   & INTRINSICS  & 0.0038 & 0.3830 & 239.85  \\
     \midrule 
       \textbf{Intel(2$\times$8$\times$1)}      & AUTOVECT    & 0.0044 & 0.3794 & 237.58  \\
     \midrule 
       \textbf{Xeon Phi(1$\times$60$\times$1)}  & KNC512      & 0.0046 & 0.2234 & 139.92  \\
     \midrule 
       \textbf{Xeon Phi(1$\times$60$\times$2)}  & NOVECT      & 0.0036 & 0.1856 & 116.26  \\
     \midrule 
       \textbf{Xeon Phi(1$\times$60$\times$4)}  & INTRINSICS  & 0.0030 & 0.1584 &  99.19  \\ 
   \bottomrule
  \end{tabular}
  \label{tab:mc_0.5M_qos_10_table} 
\end{table}     
We noted that at $50\%$ QoS none of our platform/software combinations could
satisfy the constraint in equation (\ref{eqn:constraint}). We have commented 
on this characteristic previously~\cite{Gillan:2014:VMF:2688424.2688429} explaining 
that it means only that a subset of all available options can be priced,
but not the full set. The MC kernel involves relatively expensive evaluation of 
the natural logarithm in the Box Muller transform and the exponential function 
to compute the option price.
\par
We repeated the analysis with the BT kernel, which is dominated by multiply
add operations, and report results for QoS values of $80\%$ and $40\%$ in 
tables \ref{tab:bt_4_table} - \ref{tab:bt_7_40_table}.
\input{bt_sorted_tables}

\par

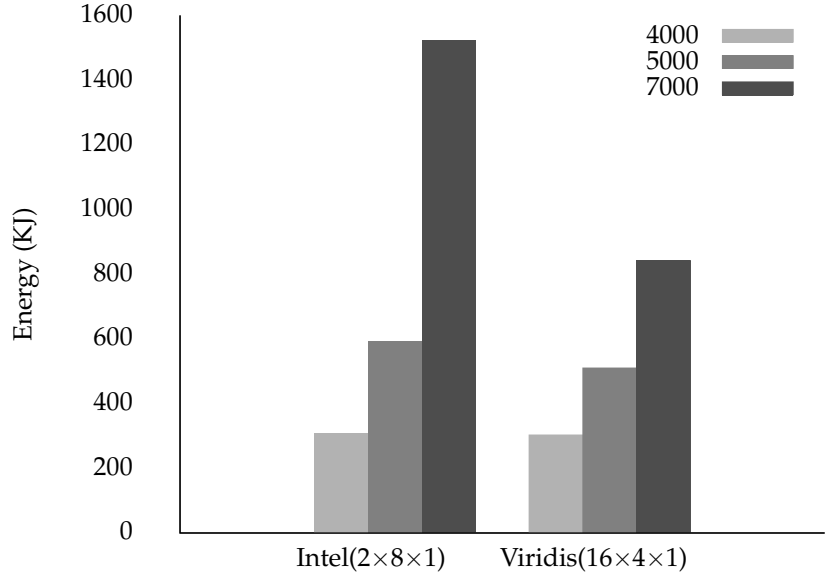
\begin{figure}[htbp]
  \centering
	\resizebox{0.65\textwidth}{!}{\input{fig5.tex}}
  \caption{BT kernel energy consumption scaling (at QoS=80\%) of Viridis(16$\times$4$\times$1) and Intel(2$\times$8$\times$1)}
  \label{fig:scaled-energy-comparison}
\end{figure}

In figure \ref{fig:scaled-energy-comparison}  we show how energy of the scaled out configurations varies with the number of points used. We are comparing Viridis(16$\times$4$\times$1) to Intel(2$\times$8$\times$1) and show how the Viridis can actually outperform Intel's Sandy Bridge while provisioning for an 80\% QoS. 

%% file: bt_sorted_tables.tex
%
%
   %
\begin{table}[htbp]
\footnotesize
\centering
\caption{BT kernel (N=4000 and QoS=80\%)}
\begin{tabular}{lrrrrrrr} 
\toprule
\multicolumn{1}{l}{\bf Platform} & 
\multicolumn{1}{c}{\bf VEC TYPE}  & 
\multicolumn{1}{c}{\bf S/Opt}  & 
\multicolumn{1}{c}{\bf J/Opt}    &
\multicolumn{1}{c}{\bf Energy(KJ)}    \\
\\

\midrule \textbf{Intel(2$\times$8$\times$1)} 
&AVX256&0.0007&0.0611&306.49\\
\midrule \textbf{Viridis(16$\times$4$\times$1)} 
&NEON128&0.0006&0.0603&302.41\\
\midrule \textbf{Intel(1$\times$8$\times$1)} 
&INTRINSICS&0.0013&0.0527&264.32\\
\midrule \textbf{Xeon Phi(1$\times$60$\times$4)} 
&INTRINSICS&0.0005&0.0131&65.88\\
\midrule \textbf{Xeon Phi(1$\times$60$\times$2)} 
&INTRINSICS&0.0004&0.0107&53.50\\
\midrule \textbf{Xeon Phi(1$\times$60$\times$1)} 
&INTRINSICS&0.0004&0.0092&46.27\\

\bottomrule
\end{tabular}
\label{tab:bt_4_table} 
\end{table}

\begin{table}[htbp]
\footnotesize
\centering
\caption{BT kernel (N=5000 and QoS=80\%)}
\begin{tabular}{lrrrrrrr} 
\toprule
\multicolumn{1}{l}{\bf Platform} & 
\multicolumn{1}{c}{\bf VEC TYPE}  & 
\multicolumn{1}{c}{\bf S/Opt}  & 
\multicolumn{1}{c}{\bf J/Opt}    &
\multicolumn{1}{c}{\bf Energy(KJ)}    \\
\\

\midrule \textbf{Intel(2$\times$8$\times$1)} 
&INTRINSICS&0.0015&0.1180&591.65\\
\midrule \textbf{Intel(1$\times$8$\times$1)} 
&INTRINSICS&0.0022&0.1017&509.69\\
\midrule \textbf{Viridis(16$\times$4$\times$1)} 
&INTRINSICS&0.0010&0.0912&457.05\\
\midrule \textbf{Xeon Phi(1$\times$60$\times$1)} 
&INTRINSICS&0.0006&0.0157&78.58\\
\midrule \textbf{Xeon Phi(1$\times$60$\times$4)} 
&INTRINSICS&0.0006&0.0152&76.23\\
\midrule \textbf{Xeon Phi(1$\times$60$\times$2)} 
&KNC512&0.0005&0.0139&69.76\\

\bottomrule
\end{tabular}
\label{tab:bt_5_table} 
\end{table}

\begin{table}[htbp]
\footnotesize
\centering
\caption{BT kernel (N=7000 and QoS=80\%)}
\begin{tabular}{lrrrrrrr} 
\toprule
\multicolumn{1}{l}{\bf Platform} & 
\multicolumn{1}{c}{\bf VEC TYPE}  & 
\multicolumn{1}{c}{\bf S/Opt}  & 
\multicolumn{1}{c}{\bf J/Opt}    &
\multicolumn{1}{c}{\bf Energy(KJ)}    \\
\\

\midrule \textbf{Intel(2$\times$8$\times$1)} 
&INTRINSICS&0.0032&0.3038&1522.85\\
\midrule \textbf{Viridis(16$\times$4$\times$1)} 
&INTRINSICS&0.0017&0.1679&841.83\\
\midrule \textbf{Xeon Phi(1$\times$60$\times$2)} 
&AUTOVECT&0.0007&0.0281&140.84\\
\midrule \textbf{Xeon Phi(1$\times$60$\times$4)} 
&INTRINSICS&0.0009&0.0275&138.02\\
\midrule \textbf{Xeon Phi(1$\times$60$\times$1)} 
&KNC512&0.0007&0.0216&108.28\\
   %
\bottomrule
\end{tabular}
\label{tab:bt_7_80_table} 
\end{table}  
   %
%
%
   %
\begin{table}[htbp]
\footnotesize
\centering
\caption{BT kernel (N=4000 and QoS=40\%)}
\begin{tabular}{lrrrrrrr} 
\toprule
\multicolumn{1}{l}{\bf Platform} & 
\multicolumn{1}{c}{\bf VEC TYPE}  & 
\multicolumn{1}{c}{\bf S/Opt}  & 
\multicolumn{1}{c}{\bf J/Opt}    &
\multicolumn{1}{c}{\bf Energy(KJ)}    \\
\\

\midrule \textbf{Intel(2$\times$8$\times$1)} 
&AVX256&0.0007&0.0611&153.24\\
\midrule \textbf{Viridis(16$\times$4$\times$1)} 
&NEON128&0.0006&0.0603&151.21\\
\midrule \textbf{Intel(1$\times$8$\times$1)} 
&INTRINSICS&0.0013&0.0527&132.16\\
\midrule \textbf{Xeon Phi(1$\times$60$\times$4)} 
&INTRINSICS&0.0005&0.0131&32.94\\
\midrule \textbf{Xeon Phi(1$\times$60$\times$2)} 
&INTRINSICS&0.0004&0.0107&26.75\\
\midrule \textbf{Xeon Phi(1$\times$60$\times$1)} 
&INTRINSICS&0.0004&0.0092&23.13\\

\bottomrule
\end{tabular}
\label{tab:bt_4_40_table} 
\end{table}

\begin{table}[htbp]
\footnotesize
\centering
\caption{BT kernel (N=5000 and QoS=40\%)}
\begin{tabular}{lrrrrrrr} 
\toprule
\multicolumn{1}{l}{\bf Platform} & 
\multicolumn{1}{c}{\bf VEC TYPE}  & 
\multicolumn{1}{c}{\bf S/Opt}  & 
\multicolumn{1}{c}{\bf J/Opt}    &
\multicolumn{1}{c}{\bf Energy(KJ)}    \\
\\

\midrule \textbf{Intel(2$\times$8$\times$1)} 
&INTRINSICS&0.0015&0.1180&295.82\\
\midrule \textbf{Intel(1$\times$8$\times$1)} 
&INTRINSICS&0.0022&0.1017&254.85\\
\midrule \textbf{Viridis(16$\times$4$\times$1)} 
&INTRINSICS&0.0010&0.0912&228.52\\
\midrule \textbf{Xeon Phi(1$\times$60$\times$1)} 
&INTRINSICS&0.0006&0.0157&39.29\\
\midrule \textbf{Xeon Phi(1$\times$60$\times$4)} 
&INTRINSICS&0.0006&0.0152&38.11\\
\midrule \textbf{Xeon Phi(1$\times$60$\times$2)} 
&KNC512&0.0005&0.0139&34.88\\

\bottomrule
\end{tabular}
\label{tab:bt_5_40_table} 
\end{table}

\begin{table}[htbp]
\footnotesize
\centering
\caption{BT kernel (N=7000 and QoS=40\%)}
\begin{tabular}{lrrrrrrr} 
\toprule
\multicolumn{1}{l}{\bf Platform} & 
\multicolumn{1}{c}{\bf VEC TYPE}  & 
\multicolumn{1}{c}{\bf S/Opt}  & 
\multicolumn{1}{c}{\bf J/Opt}    &
\multicolumn{1}{c}{\bf Energy(KJ)}    \\
\\

\midrule \textbf{Intel(2$\times$8$\times$1)} 
&INTRINSICS&0.0032&0.3038&761.42\\
\midrule \textbf{Intel(1$\times$8$\times$1)} 
&AVX256&0.0052&0.2526&632.95\\
\midrule \textbf{Viridis(16$\times$4$\times$1)} 
&INTRINSICS&0.0017&0.1679&420.92\\
\midrule \textbf{Xeon Phi(1$\times$60$\times$2)} 
&AUTOVECT&0.0007&0.0281&70.42\\
\midrule \textbf{Xeon Phi(1$\times$60$\times$4)} 
&INTRINSICS&0.0009&0.0275&69.01\\
\midrule \textbf{Xeon Phi(1$\times$60$\times$1)} 
&KNC512&0.0007&0.0216&54.14\\
   %
\bottomrule
\end{tabular}
\label{tab:bt_7_40_table} 
\end{table}  
   %
%
%

%% file: fig5.tex
\begingroup
  \makeatletter
  \providecommand\color[2][]{%
    \GenericError{(gnuplot) \space\space\space\@spaces}{%
      Package color not loaded in conjunction with
      terminal option `colourtext'%
    }{See the gnuplot documentation for explanation.%
    }{Either use 'blacktext' in gnuplot or load the package
      color.sty in LaTeX.}%
    \renewcommand\color[2][]{}%
  }%
  \providecommand\includegraphics[2][]{%
    \GenericError{(gnuplot) \space\space\space\@spaces}{%
      Package graphicx or graphics not loaded%
    }{See the gnuplot documentation for explanation.%
    }{The gnuplot epslatex terminal needs graphicx.sty or graphics.sty.}%
    \renewcommand\includegraphics[2][]{}%
  }%
  \providecommand\rotatebox[2]{#2}%
  \@ifundefined{ifGPcolor}{%
    \newif\ifGPcolor
    \GPcolorfalse
  }{}%
  \@ifundefined{ifGPblacktext}{%
    \newif\ifGPblacktext
    \GPblacktexttrue
  }{}%
  \let\gplgaddtomacro\g@addto@macro
  \gdef\gplbacktext{}%
  \gdef\gplfronttext{}%
  \makeatother
  \ifGPblacktext
    \def\colorrgb#1{}%
    \def\colorgray#1{}%
  \else
    \ifGPcolor
      \def\colorrgb#1{\color[rgb]{#1}}%
      \def\colorgray#1{\color[gray]{#1}}%
      \expandafter\def\csname LTw\endcsname{\color{white}}%
      \expandafter\def\csname LTb\endcsname{\color{black}}%
      \expandafter\def\csname LTa\endcsname{\color{black}}%
      \expandafter\def\csname LT0\endcsname{\color[rgb]{1,0,0}}%
      \expandafter\def\csname LT1\endcsname{\color[rgb]{0,1,0}}%
      \expandafter\def\csname LT2\endcsname{\color[rgb]{0,0,1}}%
      \expandafter\def\csname LT3\endcsname{\color[rgb]{1,0,1}}%
      \expandafter\def\csname LT4\endcsname{\color[rgb]{0,1,1}}%
      \expandafter\def\csname LT5\endcsname{\color[rgb]{1,1,0}}%
      \expandafter\def\csname LT6\endcsname{\color[rgb]{0,0,0}}%
      \expandafter\def\csname LT7\endcsname{\color[rgb]{1,0.3,0}}%
      \expandafter\def\csname LT8\endcsname{\color[rgb]{0.5,0.5,0.5}}%
    \else
      \def\colorrgb#1{\color{black}}%
      \def\colorgray#1{\color[gray]{#1}}%
      \expandafter\def\csname LTw\endcsname{\color{white}}%
      \expandafter\def\csname LTb\endcsname{\color{black}}%
      \expandafter\def\csname LTa\endcsname{\color{black}}%
      \expandafter\def\csname LT0\endcsname{\color{black}}%
      \expandafter\def\csname LT1\endcsname{\color{black}}%
      \expandafter\def\csname LT2\endcsname{\color{black}}%
      \expandafter\def\csname LT3\endcsname{\color{black}}%
      \expandafter\def\csname LT4\endcsname{\color{black}}%
      \expandafter\def\csname LT5\endcsname{\color{black}}%
      \expandafter\def\csname LT6\endcsname{\color{black}}%
      \expandafter\def\csname LT7\endcsname{\color{black}}%
      \expandafter\def\csname LT8\endcsname{\color{black}}%
    \fi
  \fi
  \setlength{\unitlength}{0.0500bp}%
  \begin{picture}(7200.00,5040.00)%
    \gplgaddtomacro\gplbacktext{%
      \csname LTb\endcsname%
      \put(1078,440){\makebox(0,0)[r]{\strut{} 0}}%
      \put(1078,982){\makebox(0,0)[r]{\strut{} 200}}%
      \put(1078,1524){\makebox(0,0)[r]{\strut{} 400}}%
      \put(1078,2066){\makebox(0,0)[r]{\strut{} 600}}%
      \put(1078,2608){\makebox(0,0)[r]{\strut{} 800}}%
      \put(1078,3149){\makebox(0,0)[r]{\strut{} 1000}}%
      \put(1078,3691){\makebox(0,0)[r]{\strut{} 1200}}%
      \put(1078,4233){\makebox(0,0)[r]{\strut{} 1400}}%
      \put(1078,4775){\makebox(0,0)[r]{\strut{} 1600}}%
      \put(3074,220){\makebox(0,0){\strut{}Intel(2$\times$8$\times$1)}}%
      \put(4939,220){\makebox(0,0){\strut{}Viridis(16$\times$4$\times$1)}}%
      \put(176,2607){\rotatebox{-270}{\makebox(0,0){\strut{}Energy~(KJ)}}}%
    }%
    \gplgaddtomacro\gplfronttext{%
      \csname LTb\endcsname%
      \put(5816,4602){\makebox(0,0)[r]{\strut{}4000}}%
      \csname LTb\endcsname%
      \put(5816,4382){\makebox(0,0)[r]{\strut{}5000}}%
      \csname LTb\endcsname%
      \put(5816,4162){\makebox(0,0)[r]{\strut{}7000}}%
    }%
    \gplbacktext
    \put(0,0){\includegraphics{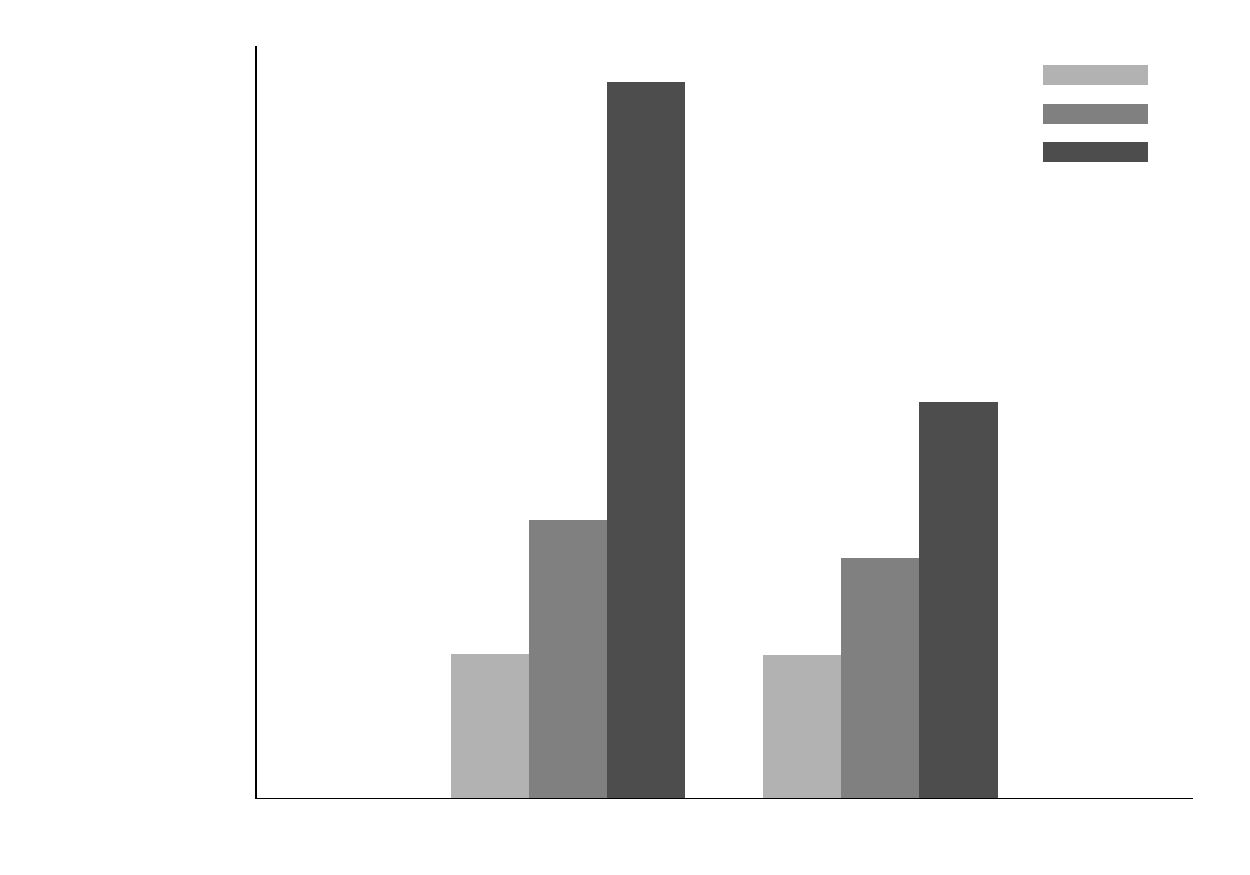}}%
    \gplfronttext
  \end{picture}%
\endgroup

%% file: discussion.tex
\section{Discussion}  \label{sec:discussion}
With a fixed number of options, achieving the $G$ constraint for a 
given QoS is inversely proportional to the S$_{\rm opt}$ metric for 
the platform and software combination. The energy consumption, therefore 
the ranking, is not only proportional to the J$_{\rm opt}$ metric but 
also depends on S$_{\rm opt}$, which determines the time needed to price 
a set of options. In our work, the top of ranking means the least 
energy consumption. 
\par
Across all the experiments, Xeon Phi is an excellent proposition 
for energy efficiency, ranking at the top. It consumes 2$\times$ up to 
an order of magnitude less energy than any other platform in any 
iso-QoS comparison. This is because Xeon Phi features a highly parallel 
and highly energy efficient manycore architecture which matches the 
parallelization and vectorization opportunities of the pricing kernels, 
especially BT. Interestingly, Xeon Phi has increasingly better energy 
efficiency compared to other platforms the higher the QoS target is and 
the more iterations the kernels performs. This means Xeon Phi energy 
efficiency scales better than in any other platform. 

\par
Viridis, scaled out to 16 nodes, ranks equivalently and up to 2$\times$ 
better than Intel across all experiments. A trend is 
visible in the BT kernel results, as the problem size increases. 
Specifically, the energy used by Viridis(16$\times$4$\times$1) rises 
more slowly than Intel the bigger the problem size. Indicatively, 
when $N=4000$, regardless the QoS target, Viridis consumes almost 
the same energy as Intel. However, when $N=7000$, Viridis uses 
approximately half the energy of the Intel configurations.

\par
Focusing on BT kernel experiment, it is interesting to note that 
details of the Xeon Phi configurations which rank at the top are different. Assuming a target QoS of $80\%$, when 
$N=4000$, the BT kernel can be served most efficiently by the Xeon Phi(1$\times$60$\times$1) 
INTRINSICS configuration. When N is increased to $5000$ this 
configuration is no longer the most energy efficient being replaced by 
the Xeon Phi(1$\times$60$\times$2) KNC512. Most interesting when moving 
to $N=7000$, the Xeon Phi(1$\times$60$\times$1) KNC512 becomes again 
the most energy efficient. Although, a higher N indicates a heavier 
computational load, the single-thread per core Xeon Phi configuration 
has better energy efficiency. This indicates that algorithmic input 
affects energy consumption in ways that are hard to predict and 
provision and we leave this investigation as future work.

\par
It is worth noting that none of the top performers involved 
compiler auto-vectorization. AUTOVECT binaries are absent from most of 
the tables because these configuration were frequently unable to 
satisfy the $G$ constraint. Observing the tables, the AUTOVECT compiler 
approach may generate the lowest S$_{\rm opt}$ metrics but this does 
not correspond necessarily to a low J$_{\rm opt}$ metric. This is 
because compiler optimizations target reducing execution time but not 
energy consumption.
\par
In addition to the use of the QoS metric to rank platforms fairly,
the graph and its analytic fit when combined with values for the S$_{\rm opt}$
and J$_{\rm opt}$ metrics allow dynamic  predictions and modeling which is of use 
to data center managers for capacity planning exercises. There are
a variety of costs involved in running a data center but simulations
of energy consumption and for a three tier (3T) configuration 
    report \cite{Kliazovich2012}
$70\%$ of the energy being consumed by the servers of 
which $43\%$, the largest single component, is from CPUs 
(modeled as running 130W). 
Economic cost models distinguish variable cost from fixed cost. 
For example, the purchase and installation of the platform represents 
the fixed cost. Our QoS metric addresses part of the so-called variable 
costs by targeting the cost of the fundamental building block of the 
service provision, namely the timely computation of option kernels. 
This allows predictive modeling of the economic option cost, which  
is associated with choosing to target the requirements of one
set of end user customers rather than another.

%% file: related_work.tex
\section{Related Work}
\label{sec:related}
Recent related work explores the performance and power consumption of servers based on low-power ARM processors~\cite{Tudor:2013:UEC:2465529.2465553,Ou:2012:ECA:2310096.2310142} suggests
that not all server workloads benefit from maximizing core counts and core frequencies, thus pinpointing opportunities for energy-efficiency optimization. Our work 
supports these findings but establishes a new metric and method for comparing servers fairly, whereby we equate the objective QoS and allow server resource scaling in our
comparisons, as opposed to equating hardware parameters such as hardware feature sizes or  core counts. 
The work of Blem et al~\cite{Blem:2013:PSR:2495252.2495491} studies the performance and power 
consumption of several ARM and Intel processors but performs head-to-head comparisons of numerous
performance and energy metrics, instead of normalizing against one key metric, which is our
approach. 
\par
Iso-metrics are common tools parallel and distributed computing. Iso-efficiency~\cite{Grama:1993:IMS:613769.613817}
in terms of sustained to theoretical maximum speedup has routinely been used to compare combinations of
parallel algorithms and architectures. Iso-energy-efficiency~\cite{Song:2011:IAS:2065358.2065695,Song:2011:IAP:2058524.2059532}
explores the influence of core scaling and frequency scaling on the energy-efficiency of algorithms and architectures.
We establish a new metric that caters to the needs of real-time analytical workloads and emerging architectures
that differ vastly in power budgets and form factors, and further establish that the new metric is more
appropriate to compare server value propositions given modern hardware diversity.  
\par
Related to our work is also prior research on improving the energy-efficiency of real-time financial workloads.
Schryver et al~\cite{Shryver11} present a methodology for efficient design of hardware accelerators for option pricing,
whereby they cap the power consumption of the accelerator and the system as a whole. 
Morales et al~\cite{Morales:2014:EFI:2616606.2616862} propose an FPGA design, programmable using OpenCL to build energy-efficient versions of binomial option pricing algorithms. They report a performance of 2,000 Options/second which is consistent  or lower than the performance attained by our Xeon Phi and scaled-out Viridis implementations, but with a power budget of 20W, which is lower than that of any of our platforms.
Hardware optimization of our workloads is beyond the scope of this paper but within the scope of our
ongoing work in the NanoStreams project~\footnote{\href{http://www.nanostreams.eu}{(http://www.nanostreams.eu)}}. The method presented in this paper fixes a workload-centric QoS metric instead of a system-centric metric, while allowing flexibility 
in tuning both system and workload parameters to meet the objective metric. 

%% file: nanostreams_project.tex
%
%
   %
\section{The NanoStreams Project} \label{sec:nanostreams}
The work reported in this paper has been carried out within the wider context of our Nanostreams
project~\footnote{\href{http://www.nanostreams.eu}{(http://www.nanostreams.eu)}}. The project bridges the 
performance gap between microservers and large servers by
enhancing microservers with application-specific, energy-efficient and programmable accelerators.
The project is building a heterogeneous microserver with a host SoC and an analytics accelerator SoC,
with a total power budget under 10 Watts, where a performance-equivalent system with state
of the art server-class processors would consume about 170 Watts.
\par
NanoStreams achieves its goals by adopting a scale-out approach
where multiple microservers and sharable accelerators are densely replicated and packaged to build systems 
with equivalent performance of large-scale servers but a dramatically smaller form factor.
A central feature of this is a co-designed software stack providing elastic and scale-free
co-execution of parallel workloads. NanoStreams uses processor-based FPGAs using dataflow processing engines
(nano-cores) and automatic C compiler generation technology to ease programming of the heterogeneous 
micro-server.  In this paper we have demonstrated that microservers are viable alternatives for low-latency,
real-time financial analytics, even if based on the now
outphased Calxeda ECX-1000 SoC and the dated Cortex A9 core. We will 
be evaluating more recent ARM-based SoCs based on 64-bit cores with GPU 
and FPGA accelerators in future work.
   %
%
%

%% file: conclusions.tex
\section{Conclusions} \label{sec:conclusions}
In this paper we have presented a mathematical formulation of an 
application-driven QoS metric for the provision of financial option 
pricing services. This metric is a function of two workload-specific but
architecture-agnostic metrics, seconds per option and Joules per option, plus several application 
parameters which define the numerical approximation computed.
Notably, our study used real stock market streaming data and captured 
the dynamic, event-driven nature of real-time financial analytics 
workloads. 
\par
Our metric facilitated direct performance comparisons between
server platforms with radically different architectural operating points and price points.
By defining a fixed QoS, a typical requirement for a service level 
agreement between a datacenter provider and the end user, we have
applied iso-QoS to rank different platforms fairly, with a repeatable workload 
using real-life and real-time data.
Our results show reveal several interesting findings: For example,
a microserver with scaled out nodes (Viridis 16$\times$4$\times$1) consumes significantly
less energy than a heavy-duty Intel Sandy Bridge server (2$\times$8$\times$1) for multiple QoS targets. When 
scaling out the number of points for computations the microserver consumes about half of the Intel
server's energy.  
\par
Our model benefits directly datacenter operators during hardware 
procurement and capacity planning exercises as it provides values which
contribute to the economic option cost of providing service to one 
or other group of end-users.
\par
Our approach creates many avenues for future research. At its most
fundamental our method allows evaluation of a QoS metric for any problem domain 
in which events, in the present case price updates, have to be processed 
by intense compute kernels, before the next event arrives. Thus the 
seconds per option metric would be replaced more generally by a seconds 
per kernel metric, similarly for the Joules per option metric.
An alternative direction of research is to incorporate the number of 
processors as a variable in the methodology and thus dynamically
provision the platforms to accommodate varying demand and a target QoS,
while attempting to minimize energy consumption. The metric can also be extended
to cater for the provisioning of heterogeneous platforms. 